\documentclass[reprint,
showpacs,
 amsmath,amssymb,
 aps,
 prl]{revtex4-1}
\usepackage{blindtext}
\usepackage{graphicx}
\usepackage[utf8]{inputenc}

\usepackage{epstopdf, epsfig}
\usepackage{amsmath}
\usepackage{float}
\usepackage{stmaryrd}
\usepackage{braket}

\usepackage{xcolor}

\usepackage{bm}

\begin{document}
\title{Angular Momentum Transport by Keplerian Turbulence in Liquid Metals}
\author{M.Vernet$^1$}
\email{marlone.vernet@phys.ens.fr}
\author{S.Fauve$^1$}
\email{stephan.fauve@phys.ens.fr}
\author{C. Gissinger$^{1,2}$}
\email{gissinger@ens.fr}

\affiliation{
$^1$Laboratoire de Physique de l’Ecole Normale Superieure, ENS, Universite PSL, CNRS, Sorbonne Universite, Universite de Paris, Paris, France\\
$^2$Institut Universitaire de France (IUF)}
\begin{abstract}
We report a laboratory study of the transport of angular momentum by a turbulent flow of an electrically conducting fluid confined in a thin disk. When the electromagnetic force applied to the liquid metal is large enough, the corresponding volume injection of angular momentum produces a turbulent flow characterized by a time-averaged Keplerian rotation rate $\overline{\Omega}\sim r^{-3/2}$. 
Two contributions to the local angular momentum transport are identified: one from the poloidal recirculation induced by the presence of boundaries, and the other from turbulent fluctuations in the bulk. The latter produces efficient angular momentum transport independent of the molecular viscosity of the fluid, and leads to Kraichnan's prediction $\text{Nu}_\Omega\propto\sqrt{\text{Ta}}$.
In this so-called ultimate regime, the experiment,  therefore, provides a configuration analogous to accretion disks, allowing the prediction of accretion rates induced by Keplerian turbulence.
\end{abstract}

\maketitle

The transport of angular momentum by turbulence is one of the most active research areas in astrophysical fluid dynamics. The best example is accretion disk theory, which  aims to understand the dynamics of thin astrophysical disks in which turbulent gas is in Keplerian rotation around a  massive central body. Observations of disks around black holes and protostars indicate enormous accretion rates which must necessarily be compensated for by a massive outward transport of angular momentum. Unfortunately, the exact mechanisms by which this transport occurs, or the nature of the turbulence in these discs remain mostly unknown~\cite{Rayleigh17,Velikhov06b, Ji06, Fromang19}. These open questions have led to a tremendous work over the past decades and different mechanisms have been proposed~\cite{Lesur05, Dubrulle05}, but the most accepted scenario is the so-called magnetorotational instability (MRI)~\cite{Balbus91}, which explains how a conducting fluid in differential rotation subjected to a magnetic field can be destabilized. Although extensively studied numerically, the experimental observation of MRI remains a major challenge for modern fluid dynamics~\cite{Sisan04,Roach12,Stefani06}, partly due to the parasitic effect of boundaries and the low saturation level of the instability in the laboratory ~\cite{Gissinger12}.

Alternatively, many studies have focused on purely hydrodynamical Taylor-Couette (TC) flow  in order to investigate the efficiency of turbulent shear flows to transport angular momentum at large kinetic Reynolds numbers~\cite{Eckhardt07, Paoletti12, Huisman12, Ji06}. A central question is whether a so-called ultimate regime, in which the angular momentum transport becomes independent of molecular viscosity $\nu$ at an arbitrary large Reynolds number $Re$, can be observed. The term  ultimate refers to the regime of thermal convection predicted by Kraichnan~\cite{Kraichnan62} in which heat transport relies entirely on convective turbulent structures, and no longer depends on the molecular diffusivity. 

However, observation of this ultimate regime is compromised by three properties specific to TC setup which have no equivalent in astrophysical disks. First, the angular momentum is injected through the rotating radial boundaries, while accretion disks are dominated by gravitation which can be regarded as a volume injection of angular momentum. 
Second, only a quasi-Keplerian rotation profile can be obtained, where the Keplerian rotation rate $\Omega=u_\theta/r\sim r^{-3/2}$ is replaced by a (presumably laminar) linearly stable flow $\Omega=A+B/r$. The transport of angular momentum then strongly depends on the exact value of the rotation ratio of the cylinders and the distance to the Rayleigh line. Third, finite size effects due to axial boundaries may, in some cases, contribute significantly to the turbulent transport ~\cite{Nordsiek15,Lopez17}. 
The first two difficulties can be partially overcome by modifying Kraichnan's theory in order to correctly describe the effect of radial boundary layers~\cite{Grossmann11}, and by rescaling the transport with an empirical function of the rotation ratio when extrapolated to Keplerian astrophysical objects~\cite{Paoletti12}. 
The role of endcaps in TC setups is more problematic, and has been the focus of a fairly active debate on the degree of turbulence generated in quasi-Keplerian flows~\cite{Ji06,Lopez17,Fromang19}. 
In addition, recent observations~\cite{Flaherty15} have suggested that turbulence in accretion disks may be weaker than expected, renewing interest in new laboratory models~\cite{Manz21,Foglizzo12, Bai21} and predictive measurements of angular momentum transport by Keplerian turbulence. In this Letter, we present a new laboratory experiment based on a radically different setup, aimed at elucidating some aspects of the turbulent transport of angular momentum and modeling accretion disks. It relies on the generation of a fully turbulent flow in an electrically conducting fluid driven by a volume Lorentz force in an axisymmetric thin disk geometry.

\begin{figure}[ht]
    \centering
    \includegraphics[width=0.4\textwidth]{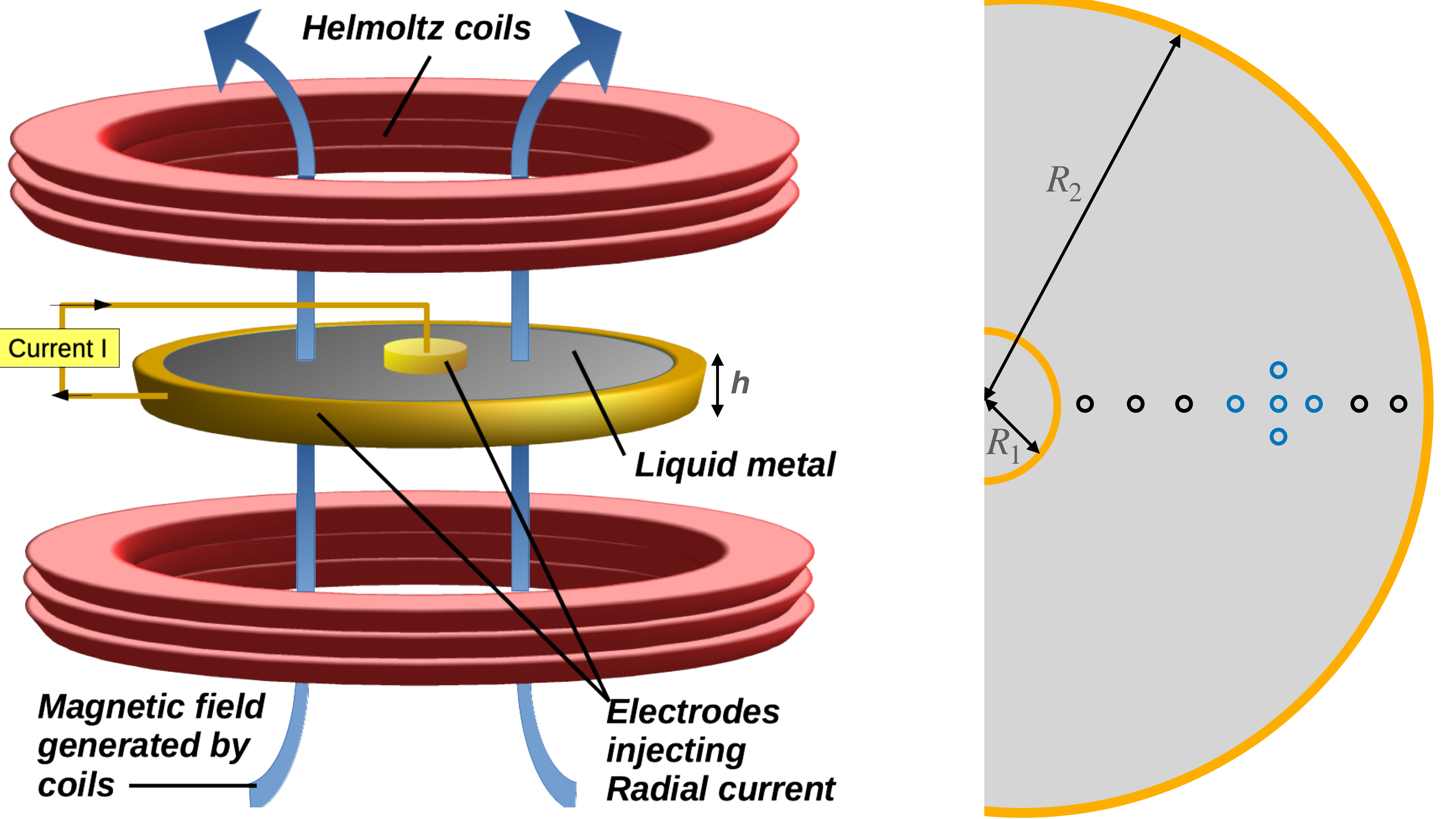}
    \caption{Left: the experimental setup is an annular cylindrical channel with inner radius $R_1=6$cm, outer radius $R_2=19$cm and height $h=1.5$cm, subjected to a radial current ($I_0=[0-3000]$ A) and a vertical magnetic field ($B_0=[0- 110]$ mT). Right: a series of potential probes extending from the top plate to midheight provide measurements of both azimuthal and radial velocity field in the midplane. The blue probes measure product $u_r\Omega$ and derivative $\partial_r\Omega$ involved in $J_\Omega$. }
    \label{fig:kepler_setup}
\end{figure}

The KEPLER experiment (see Fig \ref{fig:kepler_setup}) is an annular channel filled with liquid Galinstan and subjected to a homogeneous vertical magnetic field (up to $B_0\sim 110$mT) generated by two large Helmholtz coils. While the endcaps are Plexiglas plates, the cylinders are brass nickel-plated electrodes subjected to an electric current (up to $I_0\sim 3000$ amperes) injected radially. The resulting Lorentz force generates a turbulent flow dominated by its azimuthal component and measured from potential probes. Because the magnetic Reynolds number $\text{Rm} = \mu_0\sigma UR_1$ (with $\sigma$ the electrical conductivity and  $\mu_0$ the magnetic permeability) always remains below unity in our experiment, induction effects are negligible and the induced Maxwell stress is much smaller than the Reynolds stress.

The experimental setup as well as the main flow regimes in parameter space were presented in Ref.~\cite{Vernet21a} where it was shown that a new regime, fully turbulent, which exhibits large fluctuations and a Keplerian mean rotation profile is obtained as long as the forcing is strong enough and the disc sufficiently thin. It can be understood as resulting from the volume force balance between the Reynolds stress $\rho(u^*)^2/\delta$ (assuming a fully turbulent bulk) and the Lorentz force $I_0B_0/(2\pi r \delta)$. Here, $\delta$ is the size of the turbulent boundary layer at the endcaps,
and, similar to \cite{Kraichnan62}, the fluctuation velocity $u^*\ll \overline{u}_\theta$ is related to the mean flow by $\overline{u}_\theta/u^*=\log \text{Re}/\kappa$, where $\kappa$ is the von Karman constant and $\text{Re}=\overline{u}_\theta h/\nu$. This leads to the solution~\cite{Vernet21a}:

\begin{equation}
\overline{u}_\theta(r)=\frac{\log \text{Re}}{\kappa} \sqrt{\frac{I_0B_0}{4\pi\rho r}}   
\label{eq:kepler_profile}
\end{equation}
%Our data is very well described by this theoretical prediction, as shown by Fig.\ref{fig:kepler_profile}. Inset of Fig.\ref{fig:kepler_profile} focuses on radial profile of the angular velocity in the turbulent regime: 
where the bar denotes an average
over time. The velocity measurements reported in \textit{Vernet et al.}\cite{Vernet21a} are in very good agreement with this prediction.
 Except very close to the no-slip radial boundaries, the time-averaged flow exhibits a Keplerian rotation profile $\overline{u}_\theta\propto 1/\sqrt{r}$ over a large region of the gap, surprisingly similar to the rotation profile of an accretion disk, despite a very different origin ( the gravitation, here, being replaced by the Lorentz force).

\begin{figure}
    \centering
    \includegraphics[width=8cm]{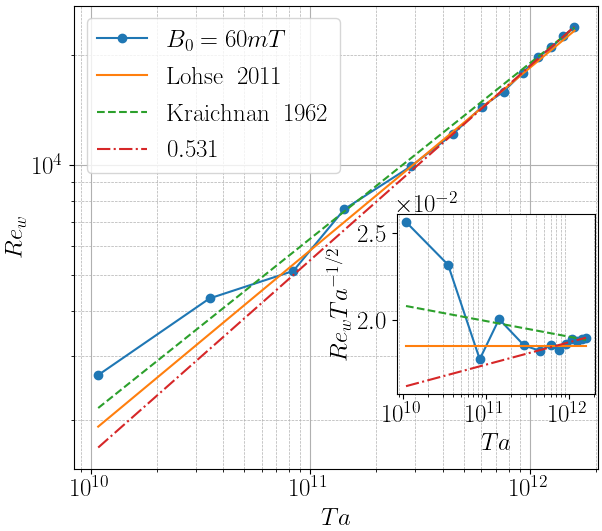}
    \caption{Wind Reynolds number $\text{Re}_w$ versus $\text{Ta}$ for an applied magnetic field $B_0=60$mT, and compared to theoretical predictions. Inset: Same, but compensated by $\sqrt{Ta}$.}
    \label{fig:Re_w_std}
\end{figure}

The turbulent transport of angular momentum reported in the present Letter shares many similarities with heat transport in thermal convection, and an exact mapping between rotational flows and Rayleigh-Benard (RB) convection can even be obtained in the limit of small radial gap and large radius~\cite{Eckhardt07, Huisman12}. The relevant quantity analogous to the heat-flux is the transverse current of azimuthal motion $\braket{J_{\Omega}} = r^3\braket{ u_r\Omega - \nu \partial_r (\Omega) }$ where  $\braket{...}$ denotes an average over time and a cylindrical surface. In the KEPLER experiment, the stationary state is given by (see Appendix):  
\begin{equation}
\partial_r\overline{J_{\Omega}}  - \frac{I_0B_0}{2\pi\rho h}r =0 
\label{eq:consJ}
\end{equation}
Taylor-Couette flows satisfy the same equation with $I_0B_0=0$, the flux $J_\Omega$ then being conserved radially as in RB convection. Here, the volume injection of angular momentum rather provides an analogy with internal or radiative heating~\cite{Kulacki72, Lepot18}, the magnetic term playing the role of a nonhomogeneous internal heating rate. Similarly, the Taylor number $\text{Ta}=\frac{4r^2\Bar{u}_{\theta}^2}{\nu^2}$ which represents the magnitude of the rotational flow is the analog of the Rayleigh number.
The level of turbulence in such rotational flows is well probed by the turbulent radial wind, which is quantified by the Reynolds number $\text{Re}_w = \frac{r u_r^*}{\nu}$ based on the fluctuations of the radial velocity $u_r^*$, here computed from the standard deviation of local measurements of $u_r(t)$. The wind Reynolds number $\text{Re}_w$ reported in Fig.\ref{fig:Re_w_std} rapidly converges to a well defined self-similar behavior at large $Ta$. Kraichnan~\cite{Kraichnan62} predicted for turbulent convection that $\text{Re}_w \propto \text{Ra}^{1/2}\text{ln}(\text{Ra})^{1/2}$, while Grossmann \& Lohse~\cite{Grossmann11} have shown that $\text{Re}_w \propto \text{Ra}^{1/2}$ (without  log correction) should be expected in turbulent flows where dissipation essentially occurs in the inertial sublayer of the turbulent boundary layers. Our data indicates an effective exponent very close to $1/2$ on almost two decades, suggesting that this last argument holds here and that the turbulent bulk supplies most of the AM transport.

 In the limit of an inductionless thin disk, the transport of angular momentum can be described by two additional dimensionless numbers (defined in the appendix): the Nusselt number $\text{Nu}_\Omega=\overline{J_\Omega}/J_{lam}$ which measures the {\it local} efficiency of the angular momentum transport compared to the laminar case, and the magnetic number $\text{H}$ measuring the strength of Lorentz force. Integration of Eq. (\ref{eq:consJ}) leads to a first relation  $\text{Nu}_\Omega \propto \text{H}/\sqrt{\text{Ta}}$, the factor of proportionality being a geometrical factor of order 1. Combining this result with the assumption of Keplerian turbulence (equation (\ref{eq:kepler_profile})) then gives:
\begin{equation}
    \text{Nu}_\Omega \propto \sqrt{\text{Ta}}\times\log^{-2}(\sqrt{\text{Ta}})
    \label{eq:NuTa2}
\end{equation}
Note that this prediction can also be recovered by a naive dimensional argument in which we suppose a simple relation $\text{Nu}_\Omega\propto \text{Ta}^\alpha \text{H}^\beta$: the requirement of a flux $J_\Omega$ independent of the molecular viscosity $\nu$ leads to a similar result $\text{Nu}_\Omega\propto \sqrt{\text{Ta}}$, sometimes referred to as the ultimate or Kraichnan regime in the literature. Turbulent RB convection leads to exponents smaller than $1/2$, generally between $0.31$~\cite{Niemela00,Ahlers09} and $0.5$~\cite{Lepot18, Roche01} depending on the experimental setup (see ~\cite{Schumacher12} for a recent review). Because of the effect of radial boundary layers, TC flows rather converge to $\text{Nu}_{\Omega}\propto\text{Ta}^{0.38}$~\cite{Gils11,Huisman12}, casting doubts on the relevance of Kraichnan's $1/2$ prediction for astrophysical objects.
\begin{figure}
    \centering
    \includegraphics[width=8cm]{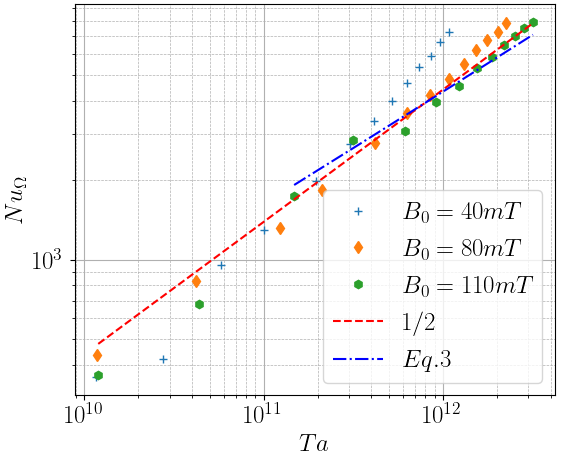}
    \caption{Nusselt number $\text{Nu}_{\Omega}$ versus Taylor number $\text{Ta}$ for typical applied magnetic fields $B_0=[40mT, 80mT, 110mT]$, compared to theoretical predictions.}
    \label{fig:Nu_Omega}
\end{figure}

For sufficiently large magnetic field, Fig.~\ref{fig:Nu_Omega} shows that  the $Nu_\Omega(Ta)$ curve follows a scaling law close to prediction (\ref{eq:NuTa2}), suggesting a turbulent transport mostly independent of the molecular viscosity. These results somehow contrast with TC flows in which the Grossman-Lohse scaling is observed, reflecting the boundary layer effects associated to the rotation of the cylinders. The KEPLER experiment exhibits an ultimate regime with smaller (although nonzero) logarithmic corrections, which naturally stems from the volume injection of angular momentum by the Lorentz force.

At smaller magnetic field however, there is a clear departure from the Kraichnan regime. This dependence of the scaling law with the magnetic field should not come as a surprise. As described in~\cite{Vernet21a}, for large enough $B_0$ the mean flow becomes two-dimensional for all scales larger than $h$. In this case, a quasi-bidimensional turbulent flow is produced, in which the poloidal recirculation is confined to thinner and thinner boundary layers as $B_0$ increases. This is one of the advantages of the present setup compared to TC flows, because a strong magnetic field decouples the bulk turbulence from the influence of axial boundaries.  The three values of $B_0$ have been chosen accordingly to this criterion, with $B_0=110$mT corresponding to a flow significantly more bi-dimensional than $B_0=40$mT.

One may expect this contribution from the poloidal recirculation to disappear with torque measurements or by averaging the flux along the zdirection. By contrast, our local measurements of $\overline{J_\Omega}$ in the mid-plane are necessarily polluted by the mean radial flow $\overline{u_r}$. This highlights the need to discriminate the contribution of this poloidal flow. To this end, we introduce the quantities:

\begin{align}
    \overline{J_\Omega}^* = \overline{J_\Omega} - r^3\overline{u_r}\overline{\Omega} ~~\text{and}~~ \text{Nu}_\Omega^* = \frac{\overline{J_\Omega}^*}{J_{lam}}
\end{align}

Here, the expression of the flux $\overline{J_\Omega^*}$ depends only on the turbulent fluctuations and is related to the Reynolds stress tensor, $\rho\overline{u_r^*\Omega^*}$, where $*$ denotes the fluctuations. Experimentally, it is obtained from direct measurements of the fluctuations, by amplifying the voltage from our potential probes through a low noise impedance matching transformer (Princeton Applied Research Model 1900).

\begin{figure}
    \centering
    \includegraphics[width=8cm]{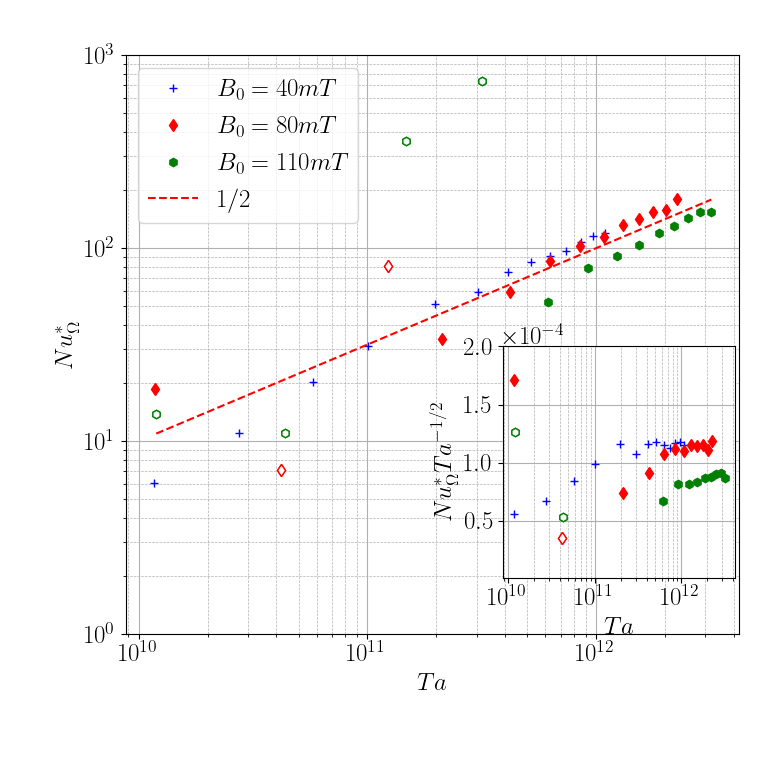}
    \caption{Fluctuation-based Nusselt number $\text{Nu}_{\Omega}^*$ versus Taylor number $\text{Ta}$ for typical applied magnetic fields $B_0=[40mT, 80mT, 110mT]$. The red dashed line corresponds to the ultimate regime $\text{Nu}_\Omega\propto\sqrt{\text{Ta}}$, only observed for large enough $B_0$ and Ta. Open symbols correspond to the oscillatory regime (see text). Inset: compensated scaling law.}
    \label{fig:Nustar_Omega}
\end{figure}

The corresponding Nusselt number $\text{Nu}_\Omega^*$ reported in Fig.\ref{fig:Nustar_Omega} can therefore be considered as a good estimate of the transport ignoring the mean poloidal recirculation $\propto\overline{u_r}\overline{\Omega}$. The values are markedly smaller than $\text{Nu}_\Omega$, but the most striking feature is the clear-cut scaling law $\text{Nu}_\Omega^*\sim \sqrt{\text{Ta}}$, now satisfied independently of the magnetic field with an exponent $1/2$ constant over two decades. This can be regarded as a measure of the angular momentum transport solely due to the turbulence in the bulk.
Note that the open symbols correspond to the low frequency oscillation reported in \cite{Vernet21a}, which does not follow Kraichnan's prediction, as expected.

Previous experiments pointed out contrasting conclusions on the role of the boundary-driven recirculation: while some studies~\cite{Paoletti12} reported large angular momentum transport and turbulence in TC flows through torque measurements, others~\cite{Ji06,schartman12} have concluded from velocity measurements in the bulk that well-controlled quasi-Keplerian flows cannot efficiently transport angular momentum. More recently, both numerical~\cite{Lopez17} and experimental~\cite{Edlund15} studies partly reconcile this contradiction, by showing that the turbulence in quasi-Keplerian flows at large $Re$ tends to recede to the boundary layers, thus leaving a relatively laminar bulk~\cite{Balbus17}. Our results offer a different perspective to this long-standing controversy: the predominance of $\overline{J_\Omega}$ over $\overline{J_\Omega^*}$ also confirms that midplane measurements, such as the torque measured in the central section in \cite{Paoletti12}, will most likely be dominated by poloidal recirculation, in our case driven by the  B\"odewadt boundary layers~\cite{Vernet21a}

As long as the electric current is kept large enough to produce a turbulent regime, increasing the magnetic field brings the flow into two-dimensional turbulence and provides an effective means of reducing this secondary flow, as illustrated by Fig.\ref{fig:Nu_Omega}. The total torque is then close (although slightly different) to Kraichnan's regime. On the other hand, in contrast to \cite{Ji06}, the bulk flow in our experiment is never laminar, as shown by Fig.\ref{fig:Nustar_Omega} in which a viscosity-free ultimate regime is observed when poloidal recirculation is ignored.

%%%%%%%%%%%%%%%%%%%%%%%%%%%%%%%%%
%%%%%%%%%%%%%%%%%%%%%%%%%%%%%%%%%%

The present experiment should not be regarded as a study of the transition to turbulence in Taylor-Couette setups or in accretion disks, which both correspond to different systems. However, it provides a turbulent flow exhibiting a mean Keplerian rotation rate and a diffusivity-free transport of angular momentum, two properties presumably satisfied by accretion disks, independently of the mechanism for the transition to turbulence.
A central question is, then, to understand to what extent the present laboratory experiment can be extrapolated to astrophysics. 
Following~\cite{Dubrulle04,Dubrulle05,Hersant05,Eckhardt07}, we define a dimensionless energy dissipation (sometimes called $\beta$) as $\text{G}^*/\text{Re}^2 = \nu^{-2}\overline{J_{\Omega}}^*/\text{Re}^2$, where $G^*$ is a dimensionless torque applied to the fluid. This quantity is related to the accretion rates $\dot{M}$ of disks by $\text{G}^*/\text{Re}^2=\dot{M}/\dot{M_0}$ (see appendix) and any deviation from $\text{G}^*\sim \text{Re}^2$ can be interpreted as an effect of the viscosity. Similarly, we have also computed the quantity $\text{G}/\text{Re}^2 = \nu^{-2}\overline{J_{\Omega}}/\text{Re}^2$ taking into account the effect of the poloidal recirculation.
\begin{figure}
    \centering
    \includegraphics[width=8cm]{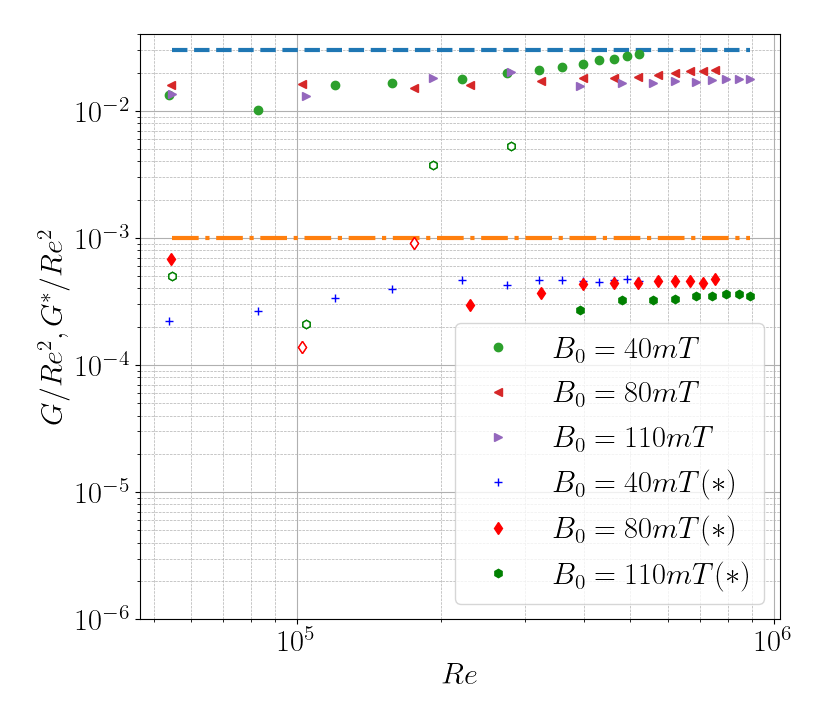}
    \caption{Dimensionless energy dissipation $\text{G}/\text{Re}^2$ and $\text{G}^*/\text{Re}^2$ versus $\text{Re}$. The dashed lines indicate the range of values obtained by \textit{Paoletti et al.}~\cite{Paoletti12} for both quasi-Keplerian and Rayleigh-unstable flows. The flow tends to an ultimate regime as the applied field is increased. (*) denotes $G^*/Re^2$ values. Open symbols correspond to the oscillatory regime (see text).}
    \label{fig:G_Re}
\end{figure}
Fig.\ref{fig:G_Re} first shows that except for $B_0=40$mT, this dimensionless energy dissipation seems to rapidly converge to a plateau, as expected for a viscosity-free regime.
 The two horizontal lines indicate the range of values given by Fig.6 of Paoletti et al~\cite{Paoletti12}, which gathers data obtained from quasi-Keplerian TC flows ($\text{G}/\text{Re}^2\sim 10^{-3}$, dash-dotted line) to counter-rotating cylinders ($\text{G}/\text{Re}^2\sim 3\times10^{-2}$, dashed line). 
 Our results for the total dissipation $\text{G}/\text{Re}^2\approx 1.8\times10^{-2}$, dominated by the poloidal recirculation, correspond to the upper bound of this previous work, related to Rayleigh-unstable Taylor-Couette flows. The dissipation due to bulk turbulence, $\text{G}^*/\text{Re}^2 \sim 4\times10^{-4}$ is much smaller and relatively close to the results obtained previously for quasi-Keplerian TC flows. This weak value, comparable to the lower bound obtained for disks around T Tauri stars~\cite{Hersant05}, may be regarded as an accurate prediction for the turbulent transport in accretion disks, for which no recirculation is present.\\
 
By exhibiting an ultimate transport of angular momentum in a turbulent and magnetized Keplerian thin disk, the KEPLER experiment provides an interesting laboratory analog of accretion disks. Naturally, these results do not aim at investigating the origin of the turbulence in astrophysical disks, but are likely to provide an interesting new constraint to the amount of angular momentum that can be transported by the turbulent fluctuations of a Keplerian disk. In this regard, it would be interesting to precisely compare these results to recent investigations claiming that some disks may be in a regime of weak turbulence~\cite{Flaherty15}.
Finally, by increasing both the size of the disk and the conductivity of the fluid, the Keplerian flow reported here may become MRI-unstable. Such an electromagnetically-driven MRI has been previously proposed as a promising setup due to the reduction of boundary effects by the presence of the Lorentz force.~\cite{Velikhov06}. It would be interesting to see the effect of the corresponding Maxwell stress on the angular momentum transport in such a large scale experiment.

\section{Acknowledgement}
This work was supported by funding from the French program
'JCJC' managed by Agence Nationale de la Recherche (Grant ANR 19-CE30-0025-01), from CEFLPRA contract 6104-1 and the Institut Universitaire de France. We thank N. Garroum, D. Courtiade and their teams for their technical assistance.

\section{Appendix}

{\bf Derivation of the equation for the current of azimuthal motion $J_\Omega$.} Magnetohydrodynamic processes are described by the combination of the Navier-Stokes and induction equations: 
\begin{align}
    \rho\partial_t\bm{u} + \rho\bm{u}\cdot\bm{\nabla}\bm{u} = -\bm{\nabla}p + \rho\nu\Delta\bm{u} + \bm{j}\times\bm{B}\\ \partial_t\bm{B} = \bm{\nabla}\times(\bm{\bm{u}\times\bm{B}}) + \frac{1}{\mu_0\sigma}\bm{\Delta B}
\end{align}
For liquid metals the magnetic Prandtl number $Pm=\mu_0\sigma\nu$ is generally very small ($\text{Pm} \sim 1.6\times 10^{-6}$ for Galinstan), leading to the so called quasi-static approximation $Pm\ll1, Rm\ll1$. In this case, the fluctuations of the magnetic field $b_i = B_0\delta_{iz} - B_i$ scale as $\sim \text{Rm}B_0$. At first order, the induction equation  reads:
\begin{align}
    B_0\partial_zu_{i} = -\eta \Delta b_{i}
    \label{eq_induc}
\end{align}
The transport of the angular velocity is related to the azimuthal component of the NS equation : 
\begin{align}
    \rho\partial_t u_{\theta} + \rho\bm{u}\cdot\bm{\nabla}u_{\theta} + \rho\frac{u_r u_{\theta}}{r} = \frac{-1}{r}\partial_{\theta}p \nonumber\\ 
    + \rho\nu\left( \partial^2_r u_{\theta} + \frac{1}{r}\partial_r u_{\theta} + \partial^2_z u_{\theta} - \frac{u_{\theta}}{r^2}  \right) \nonumber\\+\frac{1}{\mu_0}(\bm{\nabla}\times\bm{B})\times\bm{B}\cdot e_{\theta}
\end{align}
Assuming that the radial component of the current flowing across the fluid is over a fraction of the entire height $h$ is $j_r \sim I_0/2\pi r h$, one gets:
\begin{align}
    \rho\partial_t u_{\theta} + \rho\bm{u}\cdot\bm{\nabla}u_{\theta} + \rho\frac{u_r u_{\theta}}{r} = \frac{-1}{r}\partial_{\theta}(p + \frac{B^2}{2\mu_0}) \nonumber\\+ \rho\nu\left( \partial^2_r u_{\theta} + \frac{1}{r}\partial_r u_{\theta} + \partial^2_z u_{\theta} - \frac{u_{\theta}}{r^2}  \right) + \frac{I_0B_0}{2\pi rh} \nonumber\\+ \frac{1}{\mu_0}B_0\partial_zb_{\theta}
    \label{NS_theta}
\end{align}
Note that the incompressibility of the fluid gives : 
\begin{align}
    \partial_z u_z = -\frac{1}{r}\partial_r (r u_r)
    \label{div_u}
\end{align}
and the Laplacian can be rewritten as follows : 
\begin{align}
\partial_r^2 u_{\theta} + \frac{1}{r}\partial_r u_{\theta} = \frac{1}{r}\partial_r (r\partial_r u_{\theta})
\label{eq:laplace}
\end{align}

Note that for reasons of technical feasibility, all the results reported here are {\it local} measurements, carried out in the middle of the vertical gap and averaged over time only, the two-dimensional structure of the flow presumably allowing to ignore the cylindrical average. In the following, a temporal average (denoted by $\overline{X}$) is performed instead of the cylindrical average used in \cite{Eckhardt07}. Averaging NS and using the quasi-static approximation yields: 
\begin{align}
    0 =  -\overline{u_r\partial_r u_{\theta}} -\overline{u_z\partial_z u_{\theta}} - \frac{\overline{u_r u_{\theta}}}{r} \nonumber\\ -\frac{1}{r\rho}\partial_{\theta}(\overline{p + \frac{B^2}{2\mu_0}}) + \nu\left( \partial^2_r \overline{u_{\theta}} + \frac{1}{r}\partial_r \overline{u_{\theta}}  + \partial^2_z \overline{u_{\theta}} - \frac{\overline{u_{\theta}}}{r^2}  \right) \nonumber\\ - \frac{I_0B_0}{2\pi\rho rh}  - \sigma B_0^2\Delta^{-1}\partial_z^2\overline{u_{\theta}} 
\end{align}

By considering an axisymmetric time-averaged solution and making use of conditions \ref{div_u} and \ref{eq:laplace}, one can rewrite NS (after multiplying by $r^2$) :
\begin{align}
    0 =  \partial_r\left[ -r^2 \overline{u_r u_{\theta}} + \nu\left( r^3\partial_r \left(\frac{\overline{u_{\theta}}}{r}\right) \right)\right] \nonumber\\ - r^2\partial_z(\overline{u_zu_{\theta}}) + r^2\nu\partial_z^2\overline{u_{\theta}} +\frac{I_0B_0}{2\pi\rho h}r  \nonumber\\ - r^2\sigma B_0^2\Delta^{-1}\partial_z^2\overline{u_{\theta}}
    \label{eq_inter}
\end{align}
where the first term is the partial derivation of the flux of angular velocity, $J_\Omega = r^3(\overline{u_r\Omega} - \nu\partial_r\overline{\Omega})$, according to the definition given by Eckhardt \cite{Eckhardt07}. Since the flow is quasi-2D for turbulent structures larger than $h$, vertical gradients can be neglected in the bulk and equation \ref{eq_inter} becomes :
\begin{align}
    \partial_r\overline{J_{\Omega}} -  \frac{I_0B_0}{2\pi\rho h}r=0 %+ r^2\partial_z\braket{u_zu_{\theta}}_{t} 
    \label{eq_bilan}
\end{align}

The transport of angular momentum can therefore be described by only three dimensionless numbers: the Nusselt number $\text{Nu}_{\Omega}=\frac{\overline{J_{\Omega}}}{J_{lam}}$, where $J_{lam} = 2\nu r \Bar{u}_{\theta}$ is the laminar flux of angular velocity, the Taylor number $\text{Ta}=\frac{4r^2\Bar{u}_{\theta}^2}{\nu^2}$, and the magnetic number $\text{H}=\frac{I_0B_0 r}{\rho\nu^2}$. Note that equation (\ref{eq_bilan}) can  be integrated to give $\overline{J_{\Omega}} = I_0B_0 r^2/(4\pi\rho h)$, but does not directly inform about the dependence of $\text{Nu}_\Omega$ with Ta. Also, this expression should not be applied to the classical viscous-ideal regime $\Bar{u_\theta}=I_0/(4\pi r\sqrt{\rho\sigma\nu})$, as it corresponds to a limit in which  the magnetic diffusion is neglected in the induction such that all the current passes through the Hartmann boundary layers\cite{Vernet21a}, while the derivation of equation (\ref{eq_bilan}) relies on the assumption of a wider distribution of currents. 

Following \cite{Dubrulle05,Hersant05,Eckhardt07}, we also define a dimensionless torque $\text{G} = \nu^{-2}\overline{J_{\Omega}}$ and the corresponding dimensionless energy dissipation, $\text{G}/\text{Re}^2$. This quantity has a great interest as it can be related to the turbulent viscosity describing the accretion rates $\dot{M}$ of disks. More precisely, $\text{G}/\text{Re}^2=\dot{M}/\dot{M_0}$ where $\dot{M_0}$ is an effective accretion rate based on the typical sizes and rotation of the considered accretion disk~\cite{Dubrulle04,Dubrulle05,Hersant05,Paoletti12}.

\bibliographystyle{apsrev4-2} % Tell bibtex which bibliography style to use
\bibliography{biblio} % Tell bibtex which .bib file to use (this one is some example file in TexLive's file tree)

\end{document}